# Study on the formation and the decomposition of $AgN_3$ and a hypothetical compound $ReN_3$ by using density functional calculations.

G. Soto.

*Universidad Nacional Autónoma de México, Centro de Nanociencias y Nanotecnología*

*Km 107 carretera Tijuana-Ensenada, Ensenada Baja California, México.*

**Corresponding Author:** **G. Soto.**

**CNyN-UNAM**

**P.O. Box 439036, San Ysidro, CA**

**92143-9036, USA**

**Tel: +52+646+1744602, Fax: +52+646+1744603**

**E-Mail:** gerardo@cnyn.unam.mx

**Abstract**

We present a comparative study between $ReN_3$ and $AgN_3$ by using density functional theory. The $ReN_3$ is a hypothetical compound proposed by us to interpret the Re to Re interplanar spacing of thin films grown by sputtering. Both, the $AgN_3$ as the $ReN_3$, are calculated as positive enthalpy compounds. The enthalpy might give a clue about the spontaneous decomposition of the solid form, but it cannot be interpreted as a restriction of its synthesizability. As from the calculated total-energy, we discuss the route for the formation of $AgN_3$ starting from atomic species in aqueous solution. We propose that their synthesizability is conditioned by the energy of free nitrogen atoms, and the kinetics of reaction. We conclude that the intrinsic stability of a certain atomic arrangement depends only of the equilibrium of atomic forces, and not from the energy value associated with that structure.

## 1. Introduction

Predicting new solids based solely on computer calculations is one of the main challenges of materials science. Achieving this would mean a giant step forward as it would save many hours of fruitless efforts. Although there has been significant progress[1], it is still early to sing praises. Generally the prediction of materials by computer calculations seeks a global minimum of energy where it is assumed that the stable structures are to be found. In view of that, we need to review the concept of stability. When the word "stable" is applied to a material usually mean:

*(a)* A material resistant to changes of condition.
*(b)* A material that is able of maintaining its condition.

They seem to be two equivalent definitions, but are not. The stable material as defined in (*a*) is existing in a deep energy-well, where the energy barrier to any modification, either in composition or structure or state is high. Quartz is a typical example of this kind of stability. In contrast, the stable material as defined in (*b*) might be existent in a shallow energy-well. In ordinary language the materials in shallow-wells are designated as marginally stable, metastable or unstable, but this can lead to misinterpretations. The latter are stable materials because maintain their status by the internal cancellation of forces, although they may be modified by minor disturbances. Without this perturbation the state will remain unaltered forever. All materials resulting from endothermic reactions require energy input, so naturally are in a higher energy state than the reactants, and can be reversed to them easily. Thereafter they should be categorized in (*b*). The organic substances and living organism belong to the (*b*) category, this because the biologic processes need small energy barriers to function; however they are very stable in terms of their longevity and their ability to survive and adapt to different environments. The set of materials that falls within the (*b*) category is quite large. Trying to predict new solid materials seeking only for the lowest points of energy might not be a good strategy. To clarify our point we decided to make a computational study of an extreme example of a material that fall in the (*b*) category; that is, an explosive material. For this assay we use the silver azide, $AgN_3$. This is perhaps the most studied inorganic azide. The band-energy calculation has been already carried out for the $AgN_3$ compound [2]. The metal azides are characterized by the presence of the azide anion, with formula $N_3^-$. They are very sensitive explosives, and when subjected to a suitable stimulus, such as impact, heat, friction or shock, will explode. This is because the azide group retains their molecular character until a sufficient stimulus is applied to cause exothermic dissociation. The azide anion is inherently stable, but becomes unstable when subjected to small perturbations. There has been a significant interest in their structural and explosive properties since it was recognized as a component of primary explosives [3]. Simultaneously we propose a hypothetical compound between rhenium and nitrogen. We shall see that this compound is very similar to $AgN_3$ in energetic terms and composition. This $ReN_3$ compound is proposed by us to interpret the x-ray diffraction pattern and the erratic mechanic behavior of thin films growth by DC-magnetron sputter. To our understanding, there have been no predictions of energetic materials using *ab initio* calculations. If we had use a traditional energy analysis of this compound, we had to discard it because of their high energy, but by using this comparative analysis we see that this compound is viable; at least as viable as is the actual compound $AgN_3$.

**Definitions.**

The total energy, $E_t$, calculated by DFT must be translated into energies that can be interpreted by us. The $E_t$ contains the energy of cores, plus the valence electrons. To extract the chemical energy from $E_t$ we need subtract the energy of the cores, but there are many way of performing this action. Consequently we make use of the following definitions.

The cohesive energy of a $A_xB_y$ compound, $E_b^{A_xB_y}$, is defined as the transfer of energy that takes place when $x$-mol of A and $y$-mol of B, both in atomic state, reacts to form the compound. It can be calculated by DFT using the total energy as:

$$E_b^{A_xB_y} = E_t^{A_xB_y} - xE_t^{A(at)} - yE_t^{B(at)} \qquad (1)$$

Here $E_t^{A_xB_y}$ is the ground-state total energy of the $A_xB_y$ crystals. To use this formula we need to calculate $E_t^{A(at)}$ and $E_t^{B(at)}$ to the same level of accuracy as $E_t^{A_xB_y}$. These are the total energy of non-interacting A and B-atoms, usually calculated with a large cell to simulate atoms at infinite distances.

A related definition is the heat of formation, $H_f$. This is frequently expressed as an energy change, $\Delta H_f^{A_xB_y}$, that happen when *x*-mol of A and *y*-mol of B in standard conditions reacts between them to produce a mol of the compound. The calculation of $\Delta H_f$ is similar, but using the energy of the components in standard condition:

$$\Delta H_f^{A_xB_y} = E_t^{A_xB_y} - xE_t^{A(st)} - yE_t^{B(st)} \qquad (2)$$

To use the above formula we need to calculate the total energy of A and B in standard state. For example, the heat of formation of a $MeN_3$ compound, $\Delta H_f^{MeN_3}$, is defined with points of reference the energy of crystalline metal, $E_t^{Me(c)}$, and molecular nitrogen, $E(N^{mol}) = \frac{1}{2}E(N_2)$.

$$\Delta H_f^{MeN_3} = E_t^{MeN_3} - E_t^{Me(c)} - \frac{3}{2}E_t^{N_2(mol)} \qquad (3)$$

We use the expression "heat of formation" instead of "enthalpy" because the latter is defined at 273.15 K, while the calculations are done at 0 K. Often the sign of $\Delta H_f$ is source of confusion. The sign is negative for exothermic and positive for endothermic reactions, taking for granted that the reaction proceeds from the elements in standard state. We want to anticipate that the sign of $\Delta H_f$ should be never considered as an indicator of stability as defined in the (*b*)-subsection.

The concepts of cohesive energy and heat of formation were established long-ago since they are useful to study chemical reactions. For the purpose of our discussion, we want to complement with the definition of the affinity energy, $\Delta E_a$. For this we use as point of reference the energy of crystalline metal and atomic nitrogen as:

$$\Delta E_a^{Me_xN_y} = E_t^{Me_xN_y} - xE_t^{Me(c)} - yE_t^{N(at)} \qquad (4)$$

The concept of affinity is convenient to show the potential to grow the nitride once atomic nitrogen is available at the surface of a fresh metallic surface. A negative sign for $\Delta E_a$ indicates that the reaction will have tendency to happen if there is atomic nitrogen is at the surface of the metallic solid. The affinity energy was used in our earlier work of rhenium-nitrogen compounds [4], we include this definition here for those who want to compare. With a deeper look of the above definitions we can see that they are equivalent. It is easy to translate from one expression to another if the relationship between the energy of the atomic state and the energy of the standard state are known.

## 2. DFT calculation details.

For the calculations of total energies we use DFT using the Generalized Gradient Approximation (GGA) [5], as implemented in the Wien2k code [6]. The DFT with GGA has been accepted as a sufficiently good approximation to describe the structures and energies of crystalline solids. But we have to realize that it is not perfect. Often the cohesive energies are either overestimated or underestimated. The energy calibration is one of the main challenges when working with DFT since it depends on the set of basis functions [7]. Although the main objective of this work is to analyze the energetic characteristics of $ReN_3$ and $AgN_3$ compounds, we have calculated additional compounds like TiN, $Ag_3N$ and $ReN_x$, here $x$ goes from 0 to 3. The enthalpies of TiN and $AgN_3$ are reported, these can be used to adjust the enthalpies of Re-N compounds. We had as aim to use similar parameters for all calculated materials when possible. The muffin-tin radii ($r_{mt}$) were set to 2.0 au for Ti, Re and Ag atoms, and 1.0 au for N. For $AgN_3$ and $ReN_3$ we use 64 $k$-points in the irreducible part of the Brillouin zone, and 47 $k$-point for TiN. The angular momenta goes up to $l$ = 10. Optimization of volume, and when necessary, optimization of $c/a$ ($b$ = cte), $b/a$ ($c$ = cte) are performed recursively. The internal positions were found by minimizing the forces acting on all atoms. The minimum energy is established by fitting the Murnaghan equation of state to the calculated volume-energy data.

To calculate the energy of molecular nitrogen, the $N_2$ dimer was simulated by constructing a sufficiently large tetragonal unit cell (space group *P4/mmm*, $a$ = 4.0 Å, $c$ = 5.6 Å), locating $N_2$ in the vertices along the $z$-axis (N-positions (0, 0, ±0.099)) and performing the calculation for a single $k$-point at the origin of the first Brillouin zone [8]. The atomic nitrogen was calculated with a similar procedure. A cubic cell (space group $Pm\bar{3}m$, $a$ = 4.0 Å) with nitrogen at (0, 0, 0), single $k$-point and spin polarization. The atomic Ag, Re and Ti species were calculated with a cubic cell (space group $Fm\bar{3}m$, $a$ = 15.0 Å) with the metal atom at the (0, 0, 0) position. The standard state of Re (Ti) was calculated within the $P6_3/mmc$ space group, $a$ = $b$ = 2.758 (2.95), $c$ = 4.457 (4.68), $r_{\mathrm{mt}}$ = 2.0 a.u. and 42 $k$-points. The standard state of Ag was calculated within the $Fm\bar{3}m$ space group, $a$ = 4.09, $r_{\mathrm{mt}}$ = 2.0 a.u. and 47 $k$-points. The $Ag_3N$ is calculated using as model the $Cu_3N$ structure. Space group $Pm\bar{3}m$, $a$ = 4.1690 Å, Ag and N at Wyckoff positions 3$d$ and 1$a$, respectively. Table I resumes the DFT calculations of $E_t$, $E_b$, $\Delta H_f$ and $\Delta E_a$ that are used for evaluate the enthalpy of $ReN_x$.

## 3. Results

### *3.1 Energy calibration.*

As seen in Table I, the heat of formation of TiN agrees with the experimental value, however this is because to a mutual-error-canceling effect. Both the cohesive energy of crystalline titanium and molecular nitrogen are overestimated, according to refs [9,10], see Table I. With the understanding that we have carried an error, we can proceed to calculate the energies of $AgN_3$ and $Ag_3N$. In this case the cohesive energy of crystalline silver is closer as it should, in accordance with ref [10], then the mutual error cancellation do not occur. As a result, the heat of formation of both $Ag_3N$ and $AgN_3$ are underestimated, since the energy of $N^{mol}$ is overvalued by ~20 kJ mol$^{-1}$. The calculated value of $H_f$ for $AgN_3$ is 160 kJ mol$^{-1}$, while its enthalpy is reported in the range of 279 kJ mol$^{-1}$ [11] to 308 kJ mol$^{-1}$ [12]. We can correct our calculation by changing the total energy of molecular nitrogen by -0.03 Ry (by using -109.5538 Ry instead of -109.5238 Ry, a correction of -39.38 kJ mol$^{-1}$). The error in the enthalpy calculation of $AgN_3$ originates from a small deviation, ~ 0.06%, of the total energy of the $N_2$ molecule. This is an anticipated problem of Wien2k, designed to perform calculations on crystalline systems and not on molecular species; the large space between molecules cannot adequately be described by a finite set of plane waves. The buildup of error is very large for the nitrogen rich compounds; accordingly the precision of calculations for the atomic and molecular species is crucial. By using -109.5538 Ry as the energy of the molecular nitrogen, we can “correct” the enthalpy of $ReN_3$. This is just to get an idea of what should be its enthalpy, +319 kJ $mol^{-1}$, since is not completely acceptable how we did this correction. The temperature dependence cannot be used here because there is no thermodynamic data. So, we opt to use the uncorrected values henceforth. Take note that the heats of formations reported in Table IV and subsequent plots are underestimated. However $ReN_3$ enthalpy is slightly higher than the enthalpy of $AgN_3$, both are positive enthalpy compounds.

*3.2 Energy diagrams of $Ag_3N$ and $AgN_3$.*

In Figs 1 and 2 we have assembled the energy diagrams of $Ag_3N$ and $AgN_3$. The lengths of arrows are proportional to the energy of components. These are compared to their counterparts $Re_3N$ and $ReN_3$ to make an association. In these diagrams atoms at infinite distance is taken as zero chemical energy. We can see that $AgN_3$ is a positive enthalpy compound because is at higher energy than their components in standard state, while $Ag_3N$ is a negative enthalpy compound. The heats of formation for $AgN_3$ and $Ag_3N$ are in correspondence to $ReN_3$ and $Re_3N$, respectively. In some circumstances the silver azide can be synthesized, for example by treating an aqueous solution of silver nitrate with sodium azide[12]:

$$AgNO_3\,(aq) + NaN_3\,(aq) \rightarrow Ag^+ + Na^+ + (NO_3)^- + N_3^- \rightarrow AgN_3\,(s) + NaNO_3\,(aq)$$

This reaction cannot be carried out in solid state because the atomization is a crucial step. The dissociation in which the ionic $AgNO_3$ and $NaN_3$ salts separate into smaller $Ag^+$, $Na^+$, $(NO_3)^-$ and $N_3^-$ species make possible that the ion exchange occurs. Even being the $Ag_3N$ a lower energy compound, its formation is improbable here because it will require unbalance of charges at some point of reaction. The same applies to the formation of crystalline silver and molecular nitrogen. The charge balance in the aqueous solution is the limiting factor governing the above chemical reaction. This reaction result in nearly 100% of a colorless $AgN_3$ product by precipitation. However, the silver azide in solid state reacts to external perturbations, like ultraviolet light, and might explode by releasing its energy by the following solid-state route:

$$3AgN_3 + \Delta H_{released} \rightarrow 4N_2 + Ag_3N,$$

$$\text{where } \Delta H_{released} = ⅓ (4\Delta H^{N2} + \Delta H^{Ag3N} - 3\Delta H^{AgN3}) \sim -178 \text{ kJ mol}^{-1}$$

The calculate DFT energy release of $AgN_3$ is underestimated approximately by 100 kJ mol$^{-1}$, according to the analysis done previously. The important point we want to emphasize is that the chemical reactions do not always result in materials of the low energies. $AgN_3$ is a high energy compound that can be synthesized easily. The chemical reactions are thermodynamic processes in which the energy that is minimized is the total energy of process, and not the energy of an individual product.

### *3.3 The proposed structure for $ReN_3$.*

The $ReN_3$ is a hypothetical compound closely related to $AgN_3$. In Table II are the results of structure optimizations of $ReN_3$ and $AgN_3$. The cell parameters calculated by DFT for $AgN_3$ agree with the experimental values.  This gives confidence in the calculations of the $ReN_3$ structure. We arrive at the *Am2a* (40) space group after multiple iterations of local energy minimizations and force cancellations. The details of how we reach this structure are being considered for publication elsewhere. We can anticipate here that the diffraction pattern of this structure has a relative good match to the diffraction pattern of sputtered thin films, although this is not a definitive proof that this is the compound which has been deposited. Therefore it must be considered as a hypothetical structure. The proposed $ReN_3$ atomic configuration has parallel azide anions inserted between AB layers of Re atoms. It is simpler than the structure of the silver azide, which has a complex distribution of azide ions and a non-compact stacking of metal atoms. As can be see in Table III, there are N-N-N chains that act like “bonding-bridges” between the metal layers. The $N_3$-units are six-coordinated to Re atoms in $ReN_3$, while in $AgN_3$ the $N_3$-units are eight-coordinated to Ag atoms. In Table III we can see that the metal-to-nitrogen (first neighbor) distances are shorter in $ReN_3$ than in $AgN_3$, while the N-N distance increase in $ReN_3$. The bending of the $N_3$-chain makes the second neighbor distance significantly shorter in $ReN_3$ than in $AgN_3$. We are not sure that the bending of the azide anion is real in the compound $ReN_3$, but there is a lone-pair of the central nitrogen that could justify this bending. The azide compounds have interacting N-N units. Thereafter are very different in terms of their electronic structure and properties from the nitrides, which have predominantly Me-N and Me-Me electronic interactions.

*3.4 Energy calculation for $ReN_x$ compounds.*

In Table III are listed the energies calculated for rhenium-nitrogen compounds, $ReN_x$, with increasing nitrogen content, where *x* goes from 0 up to 3. The $Re_3N$ and $Re_2N$ composition are in the structures given by the Friedrich experiment [13]. The rhenium nitride, ReN, is in NiAs-like configuration as it was proposed in ref [14]. The $ReN_2$ composition was considered in our previous work [15]. The remainder compositions were taken from our earlier work [16]. The compound we want to call attention is $ReN_3$, whose structural parameters are listed in Table II. In Fig 3 we plot the trend of $E_b$ and $E_a$ as a function of composition. In this Fig we use $ReN_x$ as the molar unit, where *x* is the ratio between N to Re. Be aware in comparing the data of Table IV and Fig 3 that the energy of $ReN_{0.33}$ is ⅓ of the energy of $Re_3N$. The open symbols in plot go with to the energies of the $ReN_x$ compound. With the addition of a displacement, the $E_a$ and $E_b$ energies have the same trend. The offset is from the binding energy of a mol of $Re^c$, -780 kJ $mol^{-1}$. In the same graph the close-square symbols represent the energy if the constituents were in standard state instead of forming the compound. The mark **1** in Fig 1 corresponds to the energy of $ReN_{0.33}$. The mark **2** corresponds to ReN, and the mark **3** corresponds to $ReN_3$. The mark **4** goes with the energy of one-mol of Re in crystalline state plus the energy of three-moles of $N^{mol}$. The mark **5** is the energy released after decomposition of $ReN_3$ to $Re_3N$. We will come back to this topic later. In Fig 4 the same data is plotted, but using in this case the definition of $\Delta H_f$. The numbered tags are converted from $E_b$ to $\Delta H_f$. The interpretation of Fig 4 is straightforward. At first, when small quantities of nitrogen are introduced in the rhenium matrix, the development is to negative values for $H_f$. This means that the crystalline rhenium matrix accepts the incorporation of nitrogen by releasing energy when present in small amounts. At the $Re_3N$ composition is the most favorable point. In this case "favorable" means it takes more energy to dissociate the compound into molecular nitrogen and rhenium crystal. After this point the energy begins to increase toward positive values with further additions of nitrogen. After the 1:1 composition, the $H_f$ are positive and steadily increasing. From this point the compound can dissociate by releasing energy. The $H_f$ of $Re_3N$ is -287 kJ $mol^{-1}$, while the $H_f$ of $ReN_3$ is +201 kJ $mol^{-1}$. Like the silver azide, we propose that the rhenium azide might decompose to rhenium nitride by the following route.

$$3ReN_3 + \Delta H_{released} \rightarrow 4N_2 + Re_3N,$$

where $\Delta H_{released} = ⅓ (4\Delta H^{N2} + \Delta H^{Re3N} - 3\Delta H^{ReN3}) \sim - 297$ kJ mol$^{-1}$. (5)

The decomposition of rhenium azide is also undervalued by DFT. However we can anticipate that it will be a very energetic reaction. In Fig 4 the state after decomposition of $ReN_3$ is marked by tag **5**. The released of energy is governed by the energy differences between $Re_3N$ and $ReN_3$ as given by equation 5. The enthalpy of the compound has nothing to do here. The enthalpy is only an assisting point in the calculation of energies, but neither the formation nor the decomposition of the compound is governed by it.

## 4. Discussion

Up to this point we have anticipated by means of DFT a novel compound between rhenium and nitrogen, which is $ReN_3$. By the presence of the azide anion it should be called rhenium azide. Our calculations reveal that the $ReN_3$ is a high enthalpy compound. This proposition is different with the material given by the Friedrich experiment using high-pressure and high temperature, which is $Re_3N$ [13]. $Re_3N$ has a rather low enthalpy, and it should be a high stability point in the phase diagrams of Re-N alloys. There is an energy difference between $ReN_3$ and $Re_3N$ of nearly 300 kJ mol$^{-1}$, favorable to the formation of $Re_3N$. The question is: can the $ReN_3$ compound be synthesized? To answer that question we recur to a comparative analysis with silver nitride and silver azide. In energetic terms the developments are equivalent between the Re-N and the Ag-N systems. At low nitrogen concentrations both have compounds of low enthalpies, while at higher nitrogen concentration these have high enthalpies. However $AgN_3$ is unquestionably a synthesizable compound, despite the huge energy difference between $Ag_3N$ and $AgN_3$ of ~200 kJ mol$^{-1}$, favorable to $Ag_3N$. The $AgN_3$ is a well know explosive, upon detonation it might give $Ag_3N$ as byproduct. But in the absence of such stimulus it will retain its crystalline character. The $AgN_3$ is a stable material as it was defined in the introduction, (*b*)-subsection, given there is an internal cancellation of forces. It can be destabilized by an external event that breaks its internal equilibrium to give a violent chain reaction. We can say that the same arguments can be applied to $ReN_3$, since its structure was proposed after multiple iterations of a force cancelation algorithm. The $ReN_3$ structure, as it is proposed, have residual forces less than 2.5 mRy atom$^{-1}$, thereafter it can be said that this structure has internal equilibrium. Regardless of whether there are other structures or compositions of lower energies, under appropriate conditions this phase might be synthesized.

Now we want to talk about how new materials are predicted by *ab initio* calculations. $AgN_3$ is a higher energy point than $Ag_3N$, as determined by DFT. Obviously we do not conclude that $Ag_3N$ is synthesizable while the $AgN_3$ is not. This is because we know in advance that this will be an erroneous conclusion. However, this is the line of reasoning that is present in most of the work of prediction of new materials: always looking for lowest energy point. Occasionally there are races to see who can propose the lowest energy structure. In this line of thought, one would conclude that $Re_3N$ is a synthesizable material, while the $ReN_3$ is an unreachable composition. Is this a correct way of thinking? No, this is an over-interpretation of the basic principles of chemistry and thermodynamics. The intrinsic stability of a certain atomic arrangement depends only of the equilibrium of atomic forces. An unstable structure is the one where there is no an internal balance of forces, while in the stable structure there is balance. The energy is scalar value that is assigned to the atomic configuration in comparison to another possible configuration. The energy value is meaningful only when something is going to change from one state to another. How this structure reacts to external perturbations will give its long-term (extrinsic) stability. In the particular case of azides even a small perturbation can trigger a violent decomposition toward products of lower energy. Still the $ReN_3$ is just a hypothetical compound. We have arguable evidence that this compound was formed in sputtered films of a rhenium target in nitrogen environments. It's a shame that these results can not be published because the reviewers are biased against high enthalpy materials.

## 5. Conclusions

In this work are studied the energetic characteristics of a hypothetical rhenium azide phase. This compound belongs to a class of materials with a high amount of stored energy that can be released suddenly, to be exact, it is a potential explosive. This paper also discusses an acceptable way in which the energies of different compounds can be compared. This is important point, since the energies of compounds are often misinterpreted. The sign of the energy has nothing to do with being a synthesizable phase. However we must be aware that its synthesis can be conditioned by a suitable chemical route. For materials of lower energies may be easier to find this route.

Finally, we advise that the tactic for searching new materials should be amended. While all synthesizable structures correspond to a point of minimum energy, the energy-well could be so small that the minimum-energy search strategy may not find them. The stability of atomic structure depends from the equilibrium of atomic forces, and not from the energy value associated with that structure. As discussed earlier the silver azide is a high-energy compound, however it can be synthesized easily. From our point of view, there is a fundamental flaw in the energetic approach. This tries to describe a three-dimensional atomic arrangement by a scalar-field (energy), rather than use the more descriptive concept of vector-field (force).

## Acknowledgments

The financial support from Conacyt grant 82984 is appreciated. I appreciate the fruitful discussions with Drs. H. Tiznado and G. Moreno.

Table I.- DFT calculated total energy, cohesive, nitrogen affinity and heat of formation.

| | $E_t$ **(Ry / unit cell)** | $E_b$ **kJ mol$^{-1}$** | $\Delta E_a$ **kJ mol$^{-1}$** | $\Delta H_f$ **kJ mol$^{-1}$** |
|---|---|---|---|---|
| $N^{at}$ | -109.14867 | 0 | - | +492.502 (472) [2] |
| $N^{mol}$ | -109.52384 | -492.502 | - | 0 |
| $Ti^{at}$ | -1707.20632 | 0 | - | +545.301 (469)[1] |
| $Ti^{c}$ | -1707.62171 | -545.301 | - | 0 |
| TiN | -1817.40114 | -1373.336 | -828.035 | -335.532 (-337.65)[5] |
| $Ag^{at}$ | -10634.3387 | 0 | - | +266.077 (284)[1] |
| $Ag^{c}$ | -10634.54139 | -266.077 | - | 0 |
| $Ag_3N$ | -32013.18628 | -1340.985 | -542.753 | -50.250 |
| $AgN_3$ | -10962.99035 | -1582.712 | -1316.636 | 160.872 (279)[3], (308)[4] |
| $Ag^{(st)} + 3N^{(st)}$ | -10963.1129 | -1743.583 | - | 0 |
| $Re^{at}$ | -33435.38276 | 0 | - | +780.624 (775) [1] |
| $Re^{c}$ | -33435.97741 | -780.624 | - | 0 |
| $Re_3N$ | -100417.67499 | -3121.773 | -779.902 | -287.340 |
| $ReN_3$ | -67528.79136 | -2056.963 | -1276.338 | +201.169 |
| $Re^{(st)} + 3N^{(st)}$ | -33764.54893 | -2258.14314 | - | 0 |

[1] from ref. [10]. [2] from ref. [9]. [3] from ref. [11]. [4] from ref. [12]. [5] from ref. [17].

Table II. –Unit cells and structural parameters of $ReN_3$ and $AgN_3$.

| Compound | Figure | Structural Parameters |
|---|---|---|
| $ReN_3$ | 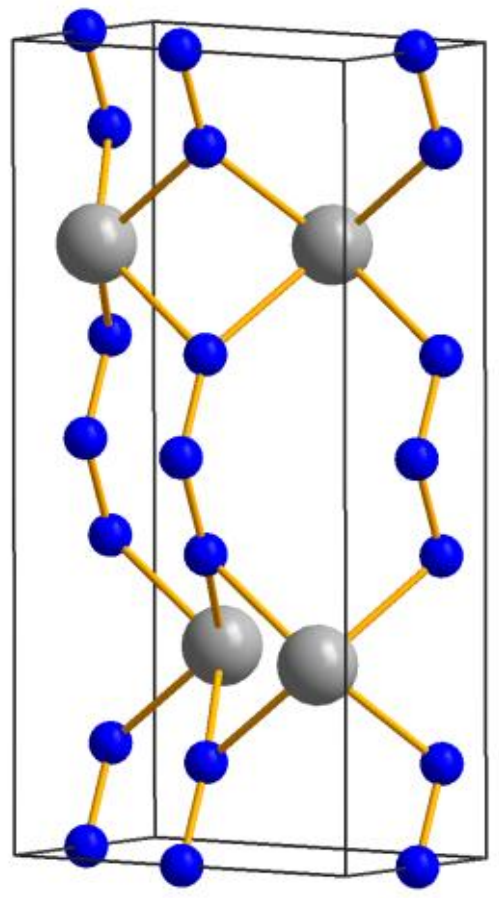 | Space Group: *Am2a* (40)<br><br>$a_0$= 10.0622, $b_0$= 5.00002,<br>$c_0$ = 2.95<br>$\alpha = \beta = \gamma = 90$<br><br>Atomic Positions:<br>Re Wyckoff 4*b*<br>(0.25, 0.66751, 0.30367)<br>N Wyckoff 8*c*<br>(0.12255, 0.00012, 0.31473)<br>N Wyckoff 4*a*<br>(0, 0, 0.48985) |
| $AgN_3$ | 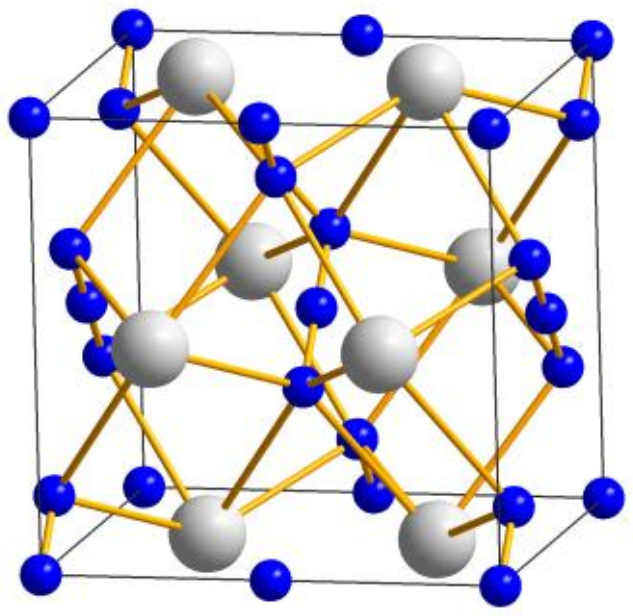 | Space Group: *Ibam* (72)<br><br>$a_0$= 5.59 (exp)[1]<br>5.6 (cal)[2]<br>$b_0$= 5.94 (exp)[1]<br>5.95 (cal)[2]<br><br>$c_0$ = 6.05 (exp)[1]<br>6.06 (cal)[2]<br><br>$\alpha = \beta = \gamma = 90$<br><br>Atomic Positions:<br>Ag Wyckoff 4*b*<br>(0.5, 0, 0.25)<br><br>N Wyckoff 4*c*<br>(0, 0, 0)<br><br>N Wyckoff 8*j*<br>(0.145, 0.145, 0)[1]<br>(0.1666, 0.1228, 0)[2] |

Table III.- Bond distances and local coordination of $N_3$ units in $ReN_3$ and $AgN_3$.

| Compound | $N_3$ local coordination | Shortest Bond Distances |
|---|---|---|
| $ReN_3$ | 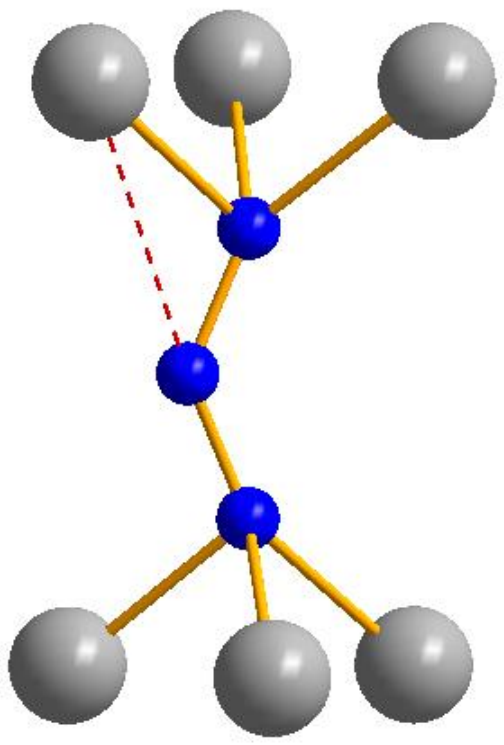 | Re-N'' : 2.1003<br>N'-N'': 1.3370<br>Re-N': 2.8083<br>Re-Re: 2.9027 |
| $AgN_3$ | 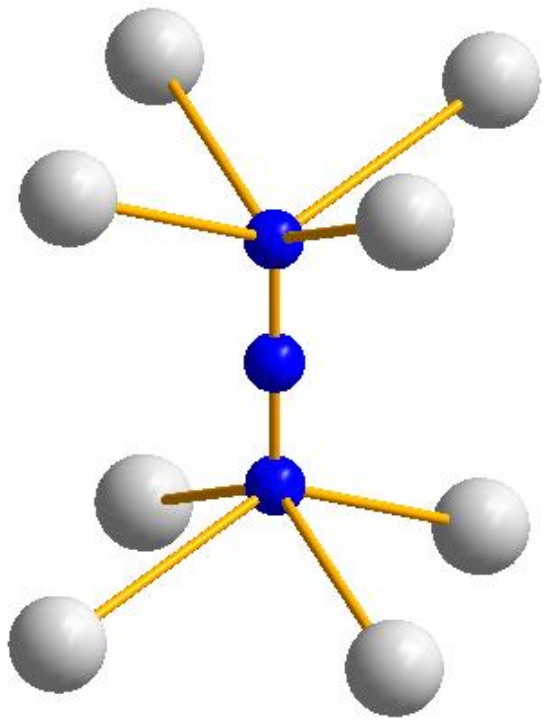 | Ag-N'' : 2.5143<br>N'-N'': 1.1863<br>Ag-N': 3.1854<br>Ag-Ag: 3.0320 |

Table IV.- Energies of $ReN_x$ compounds.

| | $\boldsymbol{x}$ | $\boldsymbol{E_t}$ **(Ry)** | $\boldsymbol{E_b}$ **kJ mol$^{-1}$** | $\boldsymbol{\Delta E_a}$ **kJ mol$^{-1}$** | $\boldsymbol{\Delta H_f}$ **kJ mol$^{-1}$** | $\boldsymbol{E_s}$ **kJ Å$^{-3}$** |
|---|---|---|---|---|---|---|
| $Re^c$ | 0 | -33435.97741 | -780.624 | 0 | 0 | 0 |
| $Re_9N$ | 0.111 | -301033.42864 | -7660.047 | -634.434 | -141.931 | -0.15061 |
| $Re_7N$ | 0.143 | -234161.82222 | -6556.161 | -1091.795 | -599.292 | -0.80847 |
| $Re_3N$ | 0.333 | -100417.67499 | -3121.773 | -779.902 | -287.340 | -5.28253 |
| $Re_2N$ | 0.5 | -133963.193 | -2208.452 | -647.204 | -154.702 | -2.21912 |
| $Re_6N_4$ | 0.667 | -201054.13715 | -6886.562 | -2202.819 | -232.810 | -0.31456 |
| $Re_8N_6$ | 0.75 | -268145.11925 | -9406.043 | -3161.053 | -206.039 | -0.20379 |
| ReN | 1 | -67091.0046 | -1274.506 | -493.883 | -1.380 | -0.02534 |
| $ReN_2$ | 2 | -67309.91897 | -1679.513 | -898.890 | +86.115 | 1.32542 |
| $ReN_3$ | 3 | -67528.79136 | -2056.961 | -1276.338 | +201.169 | 2.7094 |

# Figures

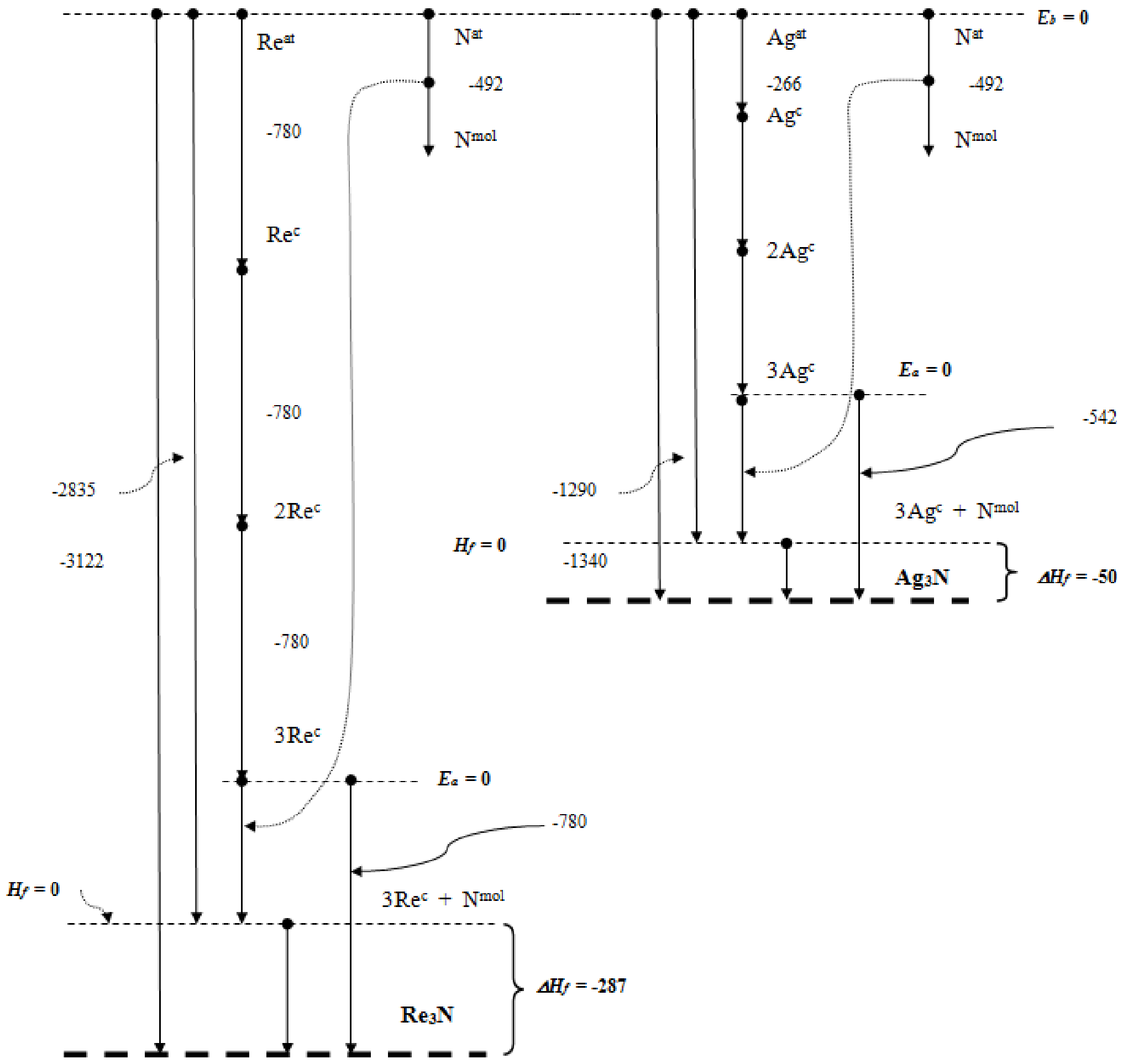

Figure 1.- Diagram of the energy formation of $Ag_3N$ ($Re_3N$) starting from: atomic species ($E_b$ =0); crystalline metal and atomic nitrogen ($E_a$ = 0); and crystalline metal and molecular nitrogen ($H_f$ = 0). All these energies are negative.

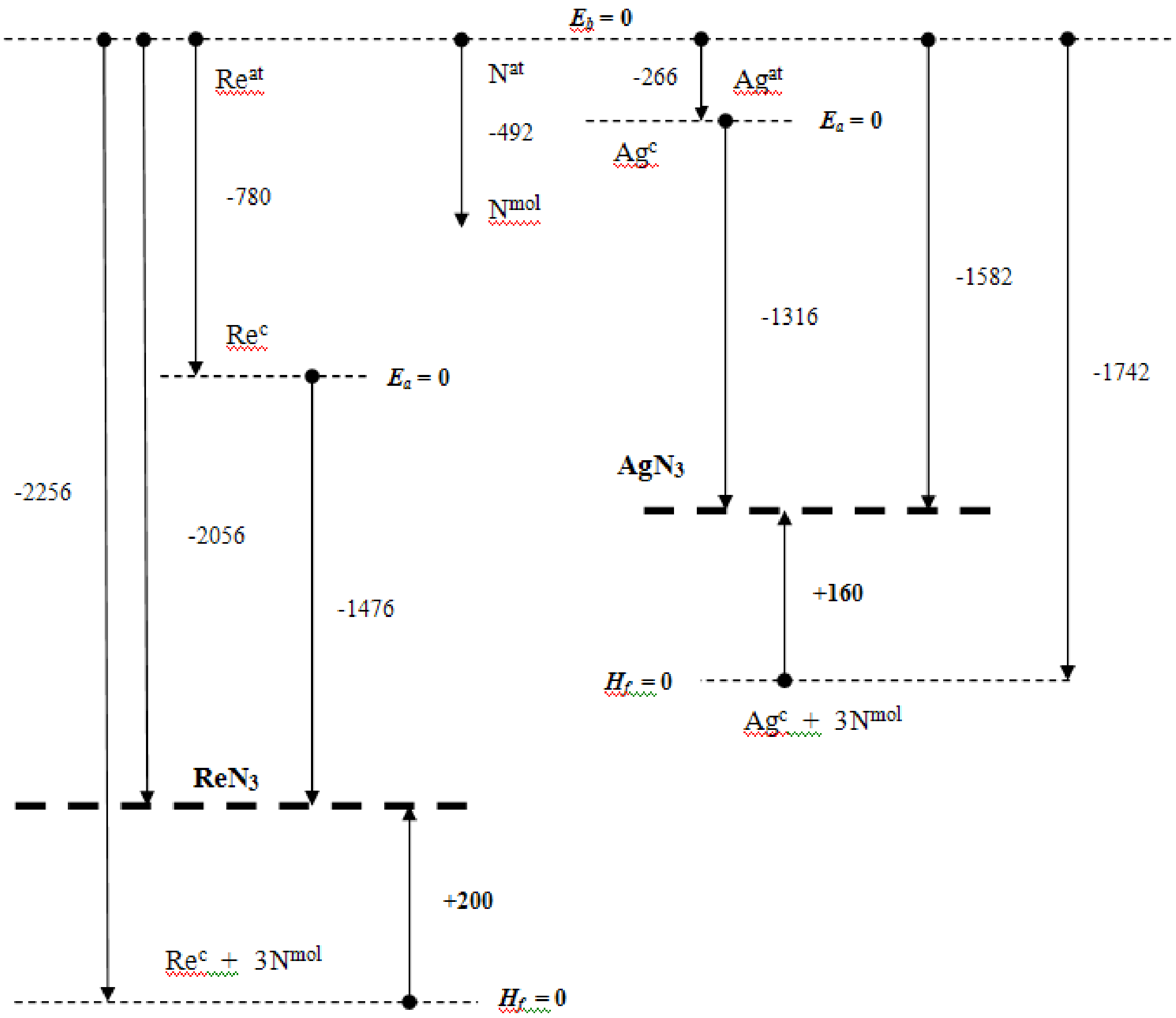

Figure 2. - Diagram of the formation energy of $ReN_3$ and $AgN_3$. In this case the heat of formations are positive, however $AgN_3$ is a real material, while $ReN_3$ is a hypothetical one.

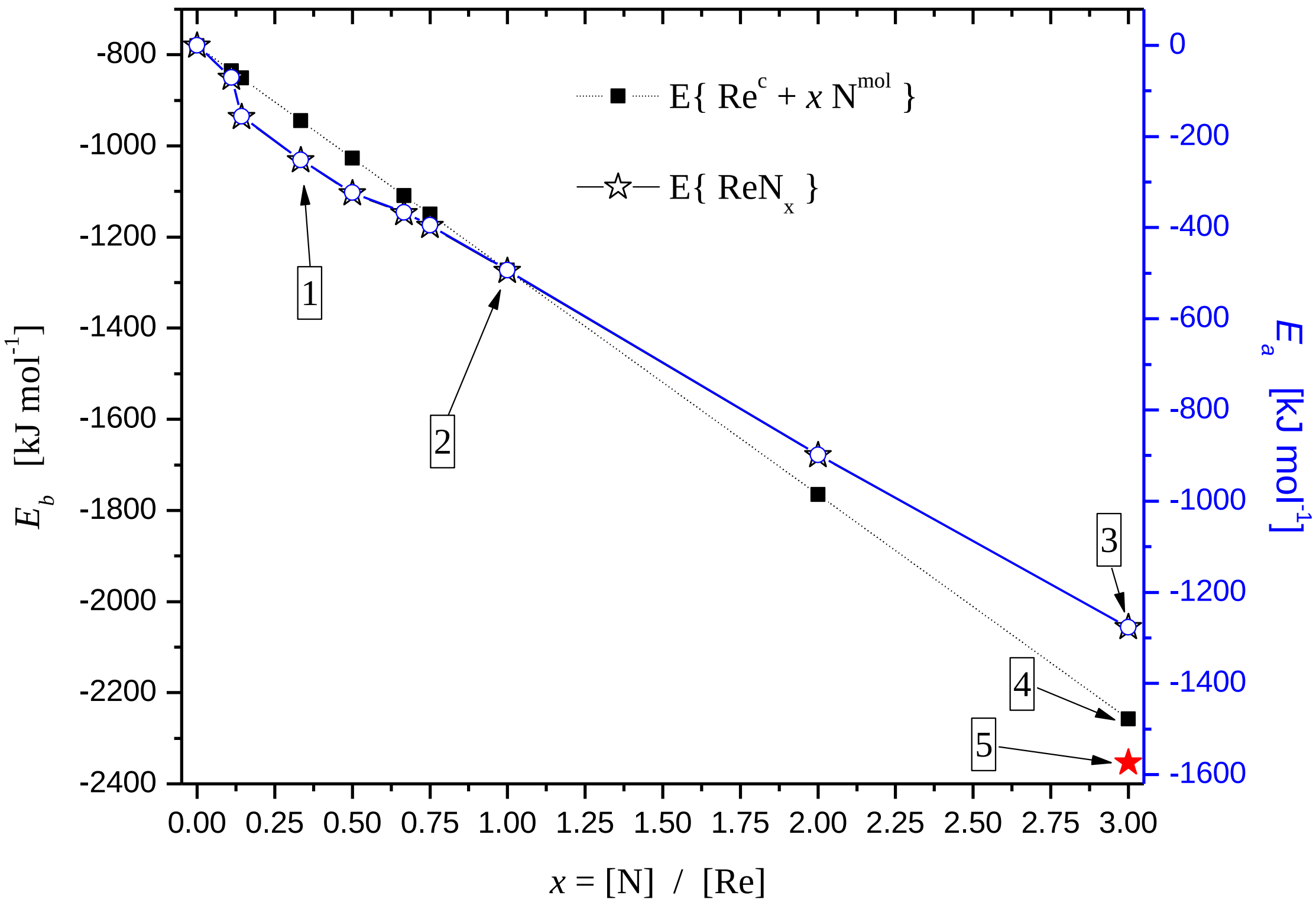

Figure 3. – Binding energy (open circle, left axis) and affinity energy (open star, right axis) as a function of nitrogen content for $ReN_x$ compounds. Also in this graph is plotted the $E_b$ energy of the standard state; this state is composed of one mol of crystalline rhenium and *x*-mol of molecular nitrogen (solid square).

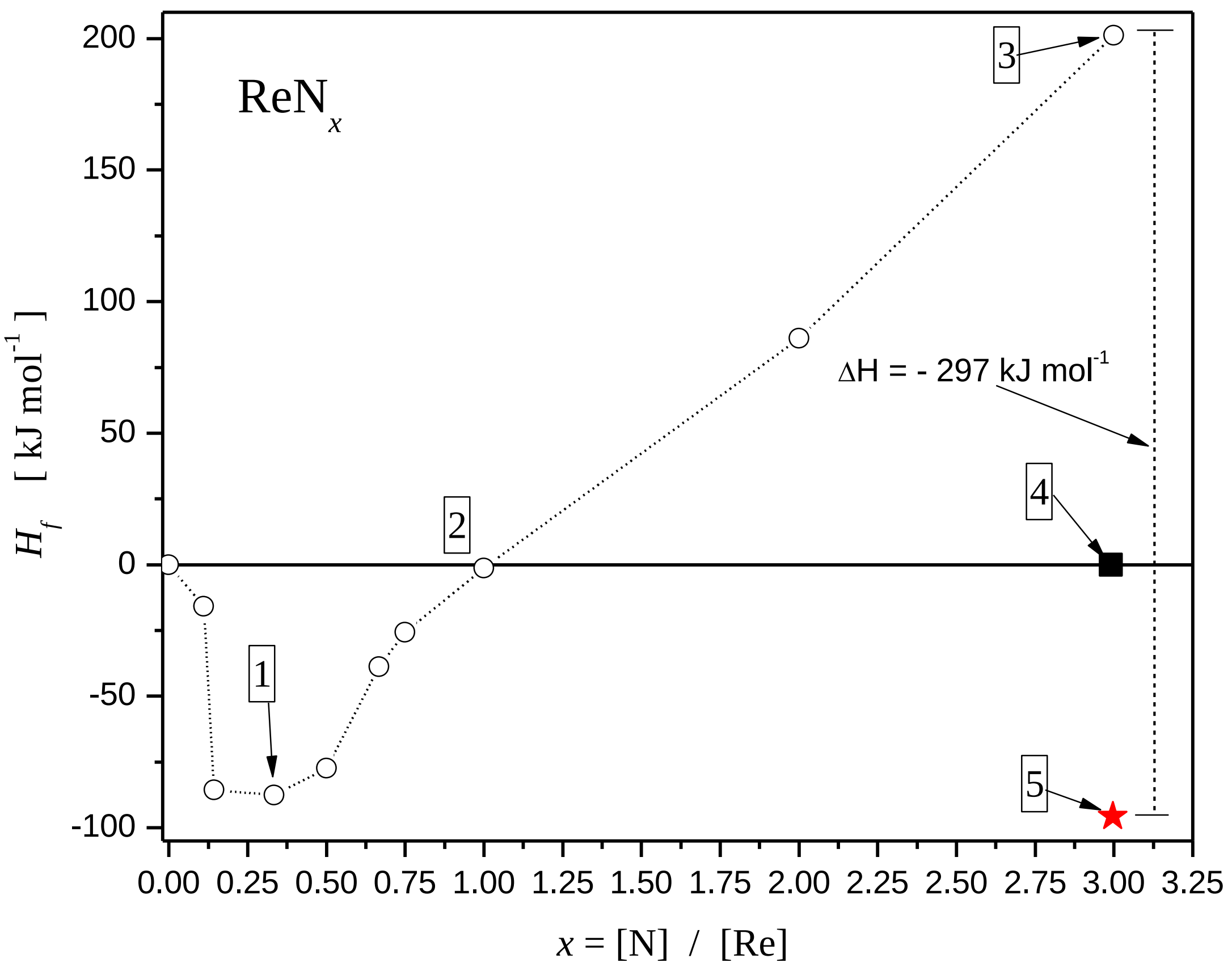

Figure 4. – Heat of formation as a function of nitrogen content for ReN$_x$ compounds. The numbered tags are described in the main text.